\documentclass[showpacs,aps,pre,twocolumn,amsmath,amssymb,superscriptaddress]{revtex4-1}
\usepackage{amsmath}           
\usepackage[latin1]{inputenc}  
\usepackage{graphicx}
\usepackage[dvipdf]{epsfig}
\usepackage{hyperref}
\usepackage{tabularx}

\begin{document}

\title{Exploiting a semi-analytic approach to study first order phase 
transitions}

\author{Carlos. E. Fiore}
\email{fiore@fisica.ufpr.br}
\author{M. G. E. da Luz}
\email{luz@fisica.ufpr.br}
\affiliation{\textit{Departamento de F\'isica,
    Universidade Federal do Paran\'a, 81531-980, Curitiba-PR, Brazil}}

\begin{abstract}

In a previous contribution, Phys. Rev. Lett {\bf 107}, 230601 (2011), we
have proposed a method to treat first order phase transitions at low 
temperatures.
It describes arbitrary order parameter through an analytical expression
$W$, which depends on few coefficients.
Such coefficients can be calculated by simulating relatively small systems, 
hence with a low computational cost.
The method determines the precise location of coexistence lines and 
arbitrary response functions (from proper derivatives of $W$).
Here we exploit and extend the approach, discussing a more 
general condition for its validity.
We show that in fact it works beyond the low $T$ limit, provided the first 
order phase transition is strong enough.
Thus, $W$ can be used even to study athermal problems, as exemplified
for a hard-core lattice gas.
We furthermore demonstrate that other relevant thermodynamic quantities, 
as entropy and energy, are also obtained from $W$. 
To clarify some important mathematical features of the method, we analyze in 
details an analytically solvable problem.
Finally, we discuss different representative models, namely, Potts, 
Bell-Lavis, and associating gas-lattice, illustrating the procedure broad 
applicability.

\end{abstract}


\maketitle

\section{Introduction}

First order phase transitions (FOPTs) occur in all sort of processes in 
nature \cite{books-first-order}, being extensively studied under different 
point of views and by a great diversity of approaches \cite{fopt-techniques}.
However, for strong FOPTs or at low temperatures, proper and reliable 
analysis still may be challenging \cite{fiore-luz1,fiore-luz2}.
This is so because in such contexts many simulation methods either can
face technical difficulties (e.g., associated to very slow convergence) 
or demand considerable computational efforts (e.g., due to the necessity 
to simulate large systems).
 
To overcome some of the above mentioned problems, in a recent short 
contribution \cite{fioreluzprl} we have proposed a general semi-analytic 
method (a not so common approach in this area \cite{pfister}) to
deal with FOPTs at low $T$'s.
The method combines simple ideas, resulting in an accurate ``combo'' protocol 
to study FOPTs.
Briefly (details in Sec. II):
(a) considering a special decomposition for the partition function at low 
$T$'s, an analytical expression $W$ to characterize FOPTs (e.g., order 
parameter) is derived;
(b) $W$ depends on some coefficients, but which can be determined through 
few numerical simulations (thus, from a computationally inexpensive 
procedure);
and 
(c) highly profiting from the analyticity of $W$ and using finite 
scale analysis for rather small systems, location of the transition 
points, order parameters behavior, and response functions (like specific 
heats and compressibilities), are obtained with good precision.

In the original paper \cite{fioreluzprl}, we have demonstrated the 
framework power by means of different examples, including the analysis of 
complicated (and often hard to simulate) Hamiltonians which describe 
diverse effects, such as water-like anomalies and ferrimagnetic-ferromagnetic
and ferromagnetic-ferromagnetic transitions.
Nevertheless, distinct important aspects of the approach were not addressed
in-depth.

Here we further explore the method, unveiling some of its mathematical 
and conceptual features.
We address the approach extent of validity, demonstrating it 
can work fine beyond the originally derived regime of applicability, i.e., 
at low $T$'s.
We propose a concrete condition (testable by simple simulations) which 
shows it can lead to good results in instances of strong FOPTs.
Besides order parameters and their derivatives (i.e., generalized 
susceptibilities), we discuss examples of other relevant thermodynamic 
quantities, like entropy and energy, that around the phase transition are 
also well described by $W$ (given that in such cases, the coefficients for 
$W$ are properly determined).

The paper is organized as the following.
The method main ideas and key expressions are summarized in Sec. II
(for completeness, full derivations are presented in the Appendix).
To clarify important technical aspects of the approach, an exactly solvable 
model \cite{huang} is discussed in Sec. III.
In Sec. IV, an extensive analysis of representative models is carried out.
The Potts model \cite{pottsreview}, displaying extreme FOPTs for large 
$q$ values (and for which the transition points are known exactly), is 
thoroughly investigated, including entropy and energy.
The method high numerical accuracy is illustrated with the Bell-Lavis model 
\cite{bell}.
Taking the associating lattice gas (ALG) model \cite{alg1,alg2} as an 
example, it is shown that the approach can be used for higher 
temperatures, provided the phase transition is sufficiently strong.
Considering a hard-core gas lattice model, it is demonstrated that even 
athermal problems can be studied with the method.
Guided by the numerical simulations and straightforward thermodynamic 
arguments, a general condition setting the approach validity is proposed 
in Sec. V.
Numerically, it is based on the calculation of the order parameter multimodal 
probability distribution at the coexistence.
Finally, remarks and conclusion are drawn in Sec. VI. 

\section{The Method}

\subsection{Main ideas and results}

The method start point \cite{fioreluzprl} is the fact that for finite 
systems at low temperatures and having $\mathcal{N}$ coexisting stable 
phases, the partition function is well described by a sum of $\mathcal{N}$ 
exponential terms, or \cite{rBoKo} ($\beta = 1/(k T)$) 
\begin{equation}
Z = \sum_{n=1}^{\mathcal{N}} \, \alpha_n \, \exp[-\beta V f_n].
\label{Z-lowT}
\end{equation}
For each phase $n$, $f_n$ is the free energy per volume $V$ and $\alpha_n$ 
the degeneracy (see also the Appendix).

Typically, relevant thermodynamic quantities have the form 
$W = -(\beta V)^{-1} (\partial/\partial \xi)\ln[Z]$, with $\xi$ an 
appropriate control parameter (e.g., chemical potential $\mu$, 
temperature, etc).
For instance, if $\xi = \mu$, the density follows directly from 
$\rho(\mu,T) = -W$ and if $\xi = \beta$, the energy per volume is 
$u = \beta W$.

As discussed in details in the Appendix, close to the transition point 
$\xi=\xi^{*}$ and considering Eq. (\ref{Z-lowT}), one finds very generally 
that $W$ can be approximated by ($y = \xi - \xi^{*}$)
\begin{equation}
W \approx (b_1 + \sum_{i=2}^{\mathcal{N}} b_i \, \exp[-a_i y])/
    (1 + \sum_{i=2}^{\mathcal{N}} c_i \, \exp[-a_i y]).
\label{eq1}
\end{equation}
The coefficients $a_i$, $b_i$ and $c_i$ are independent on the control
parameter $\xi$ and only the $a_i$'s are (linear) functions of $V$. 
In this way, at the coexistence ($y=0$), $W$ is independent on the volume 
and for all $V$ the curves $W \times \xi$ cross at $\xi=\xi^{*}$.
Therefore, $W$ in Eq. (\ref{eq1}) does not only describe order parameters, 
but it also gives the thermodynamic limit estimate for the transition point 
$\xi^{*}$.

Besides the above, two other aspects of the method, relevant for 
applications, are the following.
First, the explicit dependence of $Z$ on the free energy at low $T$'s, Eq. 
(\ref{Z-lowT}).
Close to $\xi^{*}$, it makes both $f$ and the entropy per volume $s$ 
(once $s = (u - f)/T$) also to have the same functional form of Eq. 
(\ref{eq1}).
Second, analytical derivatives of Eq. (\ref{eq1}), e.g., leading to 
specific heat, susceptibility, and order parameters which are not necessarily 
first order derivatives of the free-energy, are easily calculated.

Lastly, an important advantage of the present procedure is that Eq. 
(\ref{eq1}) is valid for any volume.
So, by considering relatively small $V$'s we can obtain the parameters 
$a$'s, $b$'s, and $c$'s with a low computational cost (see below).
It allows to describe a first order phase transition (at low temperatures) 
with a direct, accurate and numerically cheap method.
Moreover, as we are also going to show below, the approach in 
fact can be applied to broader situations than that initially assumed to 
derive Eq. (\ref{eq1}), namely, of low $T$'s.

\subsection{Numerical simulations}

Equation (\ref{eq1}) is an analytical expression to describe proper
order parameters (as well other thermodynamic functions) around the
phase transition.
Nevertheless, the coefficients $a$'s, $b$'s, and $c$'s need to be 
determined.
Although approximated expressions for these parameters do exist (Appendix), 
much better results are obtained direct from numerics.
The protocol is then: 
to use some simulation method to generate the sought thermodynamic quantity 
for different values of $\xi$; 
to compare with the corresponding curve $W$; and to determine the coefficients 
by fitting.
In particular, the actual general shape of $W$ requires only few points 
for a proper adjustment, making the numerics rather fast.
Finally, once the coefficients are known, finite size scale analysis, crossing 
determination, calculation of derivatives, etc, can all be performed 
analytically.

As a condition to choose any simulation approach to fit the coefficients 
of $W$, one should guarantee it is appropriate for the system at hands.
Then, in this work we consider the parallel tempering (PT), which is general, 
simple to use and very efficient for low and intermediate system sizes, 
even for strong FOPTs \cite{fiore2008,fiore-luz1,fiore2011}.
Such features also qualify full simulations from the PT as good benchmarks 
to test Eq. (\ref{eq1}).

For completeness, we briefly describe how to implement the PT in our 
examples (for a very detailed discussion, explaining each step in the 
method and its application to FOPT see, for instance, Ref.\cite{fiore-luz1}).
Basically, the PT (an enhanced sampling method) uses configurations from 
high to perform an ergodic walk at low $T$'s.
It simultaneously simulates a set of $R$ replicas -- in the temperature
interval $\{T_1,...,T_{R}\}$ -- by  means of a standard algorithm 
(e.g., Metropolis, cluster, etc).
When evolving any replica $i$ at a temperature $T_i$ (through an one-flip 
procedure), a given site  $k$ is chosen randomly and its state 
variable $\sigma_k'$ may change to a new value $\sigma_k''$ according to
the probability $p_i = \rm min\{1, \, \exp[- \beta_i \, \Delta {\cal H}]\}$, 
where $\Delta{\cal H} = {\cal H}(\sigma'') - {\cal H}(\sigma')$ is the 
energy change due to the transition.
Moreover, from time to time a pair of replicas (say, at $T_{i}$ and $T_{j}$ and
with microscopy configurations $\sigma'$ and $\sigma''$, respectively) can 
undergo a temperature switching, drawn from the probability
\begin{equation}                                                               
P_{T_i \leftrightarrow T_j} =                                                      
\min \{1, \, \exp[(\beta_{i} - \beta_{j})
({\cal H}(\sigma') - {\cal H}(\sigma''))] \}.                                  
\label{p-pt}                                                                   
\end{equation}
Typically, the number of replicas does not need to be very high.
For the concrete calculations in this work we set $R=12$.
We also consider adjacent and non-adjacent replica exchanges.

For the Potts model (Sec. IV.A), which presents strong discontinuous 
transition for large $q$'s, we replace the above one-flip step by the 
Wolff cluster algorithm \cite{wolff}.
In short, initially a seed site $k$ (in the state $\sigma_k$) is chosen 
at will.
Then, with probability $p = (1 - \exp[-\beta J]) \, 
\delta_{\sigma_k \, \sigma_l}$ (for $J$ the two neighbor sites interaction energy 
when both are at $\sigma_k$) $k$ is connected to each nearest neighbor site
$l$. 
This is repeated for all the new sites of the cluster until no extra site 
can be added. 
The entire cluster sites states are finally changed to a same $\sigma'$, 
randomly chosen from all the possible values for the state variable.

For the athermal hard-core lattice model we shall analyze here, very few 
modifications in the previous prescriptions are necessary.
They are explained in Sec. IV.D.

For all the examples in Section IV, we use only four points from the
simulations to determine the coefficients in Eq. (\ref{eq1}).
The resulting analytical expressions are then compared with accurate numerics 
from the PT method described above.
In many cases we also confront the present with different calculations 
in the literature to further check the efficiency of our general approach.

\section{An exactly solvable model}

We begin with an analytical example, very instructive to unveil
certain mathematical aspects of the present method.
Thus, we discuss a prototype model proposed in \cite{huang}, which 
although simple, displays all the essential aspects associated to first 
order phase transitions.
The problem grand-partition function is given by \cite{huang}
\begin{equation}
Z = (1+z)^{V}(1+z^{r V}),
\label{analytical-z}
\end{equation}
where $r$ is an arbitrary parameter and $z=\exp[\beta \, \mu]$ 
is the fugacity.
In the thermodynamic limit of $V\rightarrow \infty$, $Z$ has a real root 
at $z=1$ (i.e., $\beta \, \mu=0$), which according to the 
Yang-Lee theory \cite{yang-lee} characterizes a FOPT between two phases 
(${\mathcal N} = 2$).

Here, an appropriate order parameter is the density
\begin{equation}
\rho = \frac{1}{\beta V}\frac{\partial}{\partial \mu} \ln[Z] = 
\frac{z}{1+z}+\frac{r z^{r V}}{1+z^{r V}}.
\label{rho-analytical}
\end{equation}
Indeed, for $V \rightarrow \infty$ $\rho$ has a gap of $r$ since in such 
limit $\rho(z=1^{-}) = 1/2$ and $\rho(z=1^{+}) = 1/2 + r$.
Moreover, $\rho(z=1) = (1+r)/2$ regardless of $V$.

Now, consider a finite $\mu$.
So, low temperatures correspond to large values for $z$ (large 
$\beta \, \mu$) and a good approximation for Eq. (\ref{analytical-z}) is 
$Z \approx \exp[\beta \mu V] + \exp[\beta (r+1) \mu V]$, which obviously 
is in the general form of Eq. (\ref{Z-lowT}).
However, the phase transition takes place at $z = 1$.
For $z \approx 1$ in Eq. (\ref{analytical-z}), we can consider 
$1 + z \approx 2 \sqrt{z}$, getting  
\begin{equation}
Z \approx \exp[- \beta V f_0] + \exp[-\beta V f_r],
\label{Z-transition}
\end{equation}
with
\begin{equation}
f_0 = - \frac{1}{\beta} (\frac{\beta \mu}{2} + \ln[2]), \ \ \
f_r = - \frac{1}{\beta} ((r+\frac{1}{2}) \beta \mu + \ln[2]).
\label{f-aproximation}
\end{equation}
Notice that again $Z$ takes the form of Eq. (\ref{Z-lowT}).
Hence, although being a particular model, it shows that Eq. 
(\ref{Z-lowT}) and so Eq. ({\ref{eq1}) can hold true in more general 
instances than just that of low $T$'s (see the discussion in Sec. V).

As mentioned in Sec. II, the density is given by $\rho = -W$ with 
$\xi = \mu$.
Thus, using the analytical relations for the coefficients of Eq. 
(\ref{eq1}) in the Appendix, one finds that for the present case:
$c_2 = 1$, $b_1 = -1/2$, $b_2 = -(r+1/2)$, and $a_2 = r \beta V$.
Therefore, close to the phase transition point it follows from 
Eq. (\ref{eq1}) an approximation for $\rho$, or
\begin{equation}
\rho_{z \sim 1} = \frac{1}{2} +
\frac{r z^{r V}}{1 + z^{r V}}, 
\label{rho-aprox}
\end{equation}
which yields the correct limits once $\rho_{z \sim 1}(z=1) = \rho(z=1)$ does 
not depend on $V$ and 
$\rho_{z \sim 1}(1^+) = - b_2/c_2 = \rho(1^+)$ and
$\rho_{z \sim 1}(1^-) = - b_1/c_1 = \rho(1^-)$
for $V \rightarrow \infty$.

\begin{figure}[top]
\centerline{\psfig{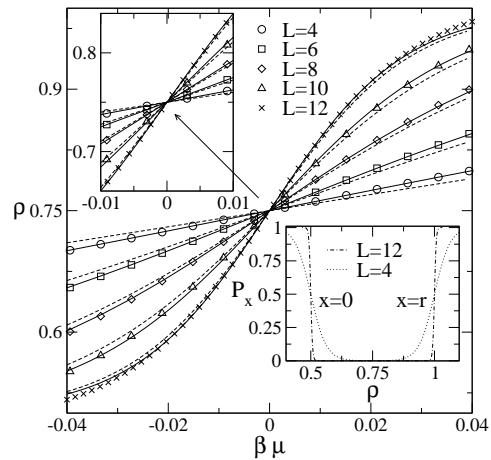}} 
\caption{$\rho$ vs. $\beta \, \mu$ for distinct $V = L \times L$ and 
$r = 1/2$.
The left inset is a blow up around the transition point, at which 
$\rho=3/4$. 
Symbols correspond to the exact $\rho$ and dashed lines to Eq. 
(\ref{rho-aprox}).
The continuous lines are for the density given by the general Eq. 
(\ref{eq1}), with the parameters $c_2$ and $b$'s the same ones used in Eq. 
(\ref{rho-aprox}), but $a_2/\beta$ from a best numerical fitting.
For two $V$ values and $r=1/2$, $P_0(\rho)$ and $P_r(\rho)$ are plotted 
together in the right inset.
Since $P_0$ ($P_r$) is practically zero for $\rho \geq 3/4$ 
($\rho \leq 3/4$), the overlap between the two distributions is
negligible.}
\label{fig1}
\end{figure}

In Fig. \ref{fig1} we show the exact $\rho$ vs. $\beta \, \mu$ for 
different volumes $V = L \times L$ and $r=1/2$. 
We compare such curves with Eq. (\ref{rho-aprox}) and also with Eq. 
(\ref{eq1}) for the $c_2$, $b_1$, and $b_2$ as above and $a_2/\beta$ given by 
the best numerical fitting to Eq. (\ref{rho-analytical}) (see Table I).
For all the volumes, we observe a very good agreement between the 
exact $\rho$ and the approximations, specially in the case where
$a_2/\beta$ is fitted. 
This latter demonstrates that the general Eq. (\ref{eq1}) -- with the
coefficients obtained from numerical simulations -- is an accurate procedure 
to calculate FOPTs order parameters.

As a final analysis, we recall that the probability to be in phase 
$x$ (for $x = 0 \ \mbox{or} \ r$) is $P_x = w_x/Z$, with $Z = w_0 + w_r$ 
and $w_x$ the proper weight of phase $x$.
Also, in the thermodynamic limit the term $r z^{r V}/(1 + z^{r V})$ in the
exact $\rho$, Eq. (\ref{rho-analytical}), is just the Heaviside function 
$\Theta(z-1)$ times the parameter $r$.
Hence, it is the term $z/(1+z)$ in Eq. (\ref{rho-analytical}) that 
actually gives the density variation with respect to the fugacity.
Thus, considering Eqs. (\ref{Z-transition})-(\ref{f-aproximation})
close to the transition point -- but still at the phase $x$ -- the 
probability and density, as function of $z$, read
\begin{equation}
P_x \approx \frac{1}{Z} \exp[-\beta V f_x] = \frac{z^{x V}}{1 + z^{r V}},
\ \ \ 
\rho_x \approx x + \frac{z}{1+z}.
\label{p-rho}
\end{equation}

By isolating $z$ in Eq. (\ref{p-rho}) one obtains $P_x(\rho_x)$, which gives 
(around $z \approx 1$) the probability to be in phase $x$ with the density 
value $\rho_x$.
Plots for two distinct $V$'s and $r=1/2$ are shown in the right inset 
of Fig. 1.
It is interesting to observe that although $P_x(\rho_x)$ becomes broader 
for lower $V$'s, even for a so small volume of $V = 4 \times 4$ the density 
distributions of the two phases do not overlap if $r=1/2$.
On the other hand, by decreasing $r$, $P_0$ and $P_r$ start to intersect 
each other.
But $r$ measures the jump in the order parameter, consequently how 
strong is the phase transition. 
Thus, a weaker phase transition leads to a larger overlap between the 
probability distributions of the order parameter close to the transition
point.
As it will be discussed in Sec. V, this result illustrates a general and 
important fact to set the method validity.

\begin{table}[t]
\label{tableI}
\caption{The values of $a_2/\beta$ used in Figure 1.}
\begin{tabular}{l l l}
\hline
\hline
$L$, where $V = L \times L$  \ \ & $a_2/\beta = r V$  \ & $a_2/\beta$ 
(numerical fitting)
\\ \hline
$L=4$
&
8
&
10.016233
\\
$L=6$
&
18
&
20.073368
\\
$L=8$
&
32
&
34.164822
\\
$L=12$
&
72
&
75.295494
\\
\hline \hline
\end{tabular}
\end{table}

\section{Numerical Applications}

\subsection{Potts model}

Widely studied in statistical mechanics, both analytically
\cite{baxter,kim,baxter81,borgs2} and numerically \cite{janke,ch-wang,wang2},
the Potts model is a fine test for the present method, 
specially given its transition points are exactly known
\cite{pottsreview}.

Consider each site of a regular lattice associated to a spin 
variable $\sigma$, which assume the values $0, 1, \ldots, q-1$.
If two adjacent sites have different (same) spin, their interaction 
energy is null ($-J$).
Thus, the Hamiltonian reads
\begin{equation}
{\cal H} = -J \sum_{(i,j)} \delta_{\sigma_i \, \sigma_j}.
\end{equation}
In the limit of very low temperatures, the system is constrained to
an ordered phase, becoming disordered as $T$ increases. 
For any $q$, the order-disorder transition takes place at 
$T_{c} = 1/\ln[1+\sqrt{q}]$, that in two dimensions is second-order for 
$q \le 4$ and first-order for $q \geq 5$.
An appropriate order parameter is 
\begin{equation}
\phi = \frac{q (V_{max} / V)-1}{q-1},
\label{order-potts}
\end{equation}
where $V_{max}$ is the volume occupied by the spins in a state $\sigma$ of
largest population and $V = L^2$ (in 2D) is the total volume 
\cite{baxter,kim,baxter81}. 
  
\begin{figure}[top]
\centerline{\psfig{figure=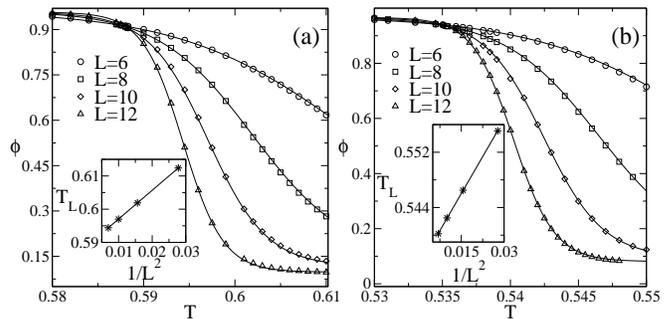,scale=0.3}} 
\caption{The Potts model order parameter $\phi$ versus $T$ for 
(a) $q=20$ and (b) $q=30$, and different system sizes $L$. 
Continuous lines are from Eq. (\ref{eq1}).
The curves cross at $T_0=0.5883(3)$ in (a) and at $T_0=0.5352(1)$ in (b). 
The insets show $T_L$ versus $1/L^{2}$, where $T_L$ is the temperature at 
the peak of $\chi = - (\partial/\partial T) \phi$.}
\label{fig2}
\end{figure}

For the numerics we set $q=20$ and $q=30$, values which characterize
strong first-order phase transitions. 
In Fig. \ref{fig2} we show the order parameter, Eq. (\ref{order-potts}), 
as function of $T$.
We clearly see that in all cases $\phi$ is well described by Eq. 
(\ref{eq1}) (through proper parameters fitting).
Moreover, in Fig. \ref{fig2} the crossing points are at $T_0=0.5883(3)$ 
for $q=20$ and at $T_0=0.5352(1)$ for $q=30$.
Such values are corroborated by finite size scaling obtained from $T_{L}$ 
versus $1/L^{2}$, with $T_{L}$ the temperature at the peaks of the response 
function $\chi = -(\partial/\partial T) \phi$ \cite{rBoKo,ch-wang}.
Indeed, extrapolating the plots of $T_L$ versus $1/L^2$ (insets of Fig. 
\ref{fig2}), we find $T_0=0.5881(1)$ for $q=20$ and $T_0=0.5353(3)$ 
for $q=30$.
The estimations are in excellent agreement with the exact values
$1/\ln[1+\sqrt{20}] = 0.588349(\ldots)$ and 
$1/\ln[1+\sqrt{30}] = 0.535248(\ldots)$.

As mentioned in the Sec. II.A, the mean energy and entropy, $u$ and 
$s$, per site are also described by an expression in the form of Eq. 
(\ref{eq1}).
In Fig. \ref{fig3} (Fig. \ref{fig4}) we display the results for 
$q=20$ ($q=30$).
Since entropy is not directly computed from standard thermodynamic 
averages, here we consider an indirect procedure, based on the transfer 
matrix method (for details refer, e.g., to \cite{sauerwein,fiore-luz2}), 
to make the simulations and fit the coefficients of Eq. (\ref{eq1}) in the 
case of $s$.
We note that once more Eq. (\ref{eq1}) indeed does describe quite well the 
relevant thermodynamic quantities and the crossing point (for a similar 
analysis for $u$ and $s$, but considering a different approach, see 
Ref.\cite{wang2}).
We also can perform finite size scaling from the specific heat 
$c_V= (\partial/\partial T) u$, plotting $T_{L}$ versus $1/L^2$, 
for the $T_{L}$'s the peak positions of the curves $c_V(T)$ 
for distinct $L$'s \cite{rBoKo,ch-wang} (insets of Figs.  
\ref{fig3} and \ref{fig4}). 
From the $u$'s crossing and the $T_L$ extrapolation we find,
respectively, $T_0 = 0.5882(2)$ and $T_0 = 0.5882(1)$ for $q=20$ and
$T_0 = 0.5353(4)$ and $T_0 = 0.5351(1)$ for $q=30$, again consistent
with the previous results.

\begin{figure}[top]
\centerline{\psfig{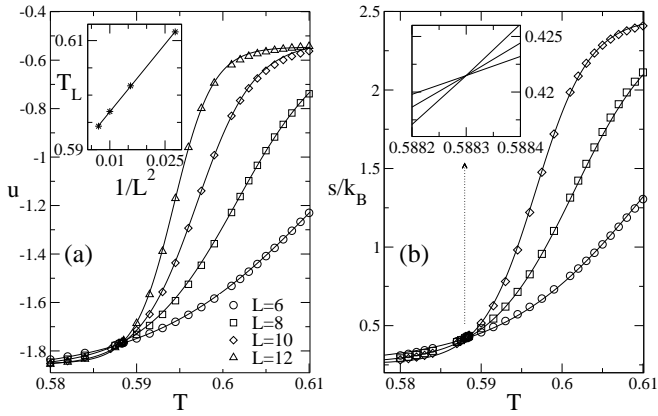}} 
\caption{The Potts model with $q=20$ and different $L$'s.
The continuous lines are from Eq. (\ref{eq1}).
(a) $u$ versus $T$, with the crossing at $T=0.5882(2)$.  
The inset displays $T_L$ versus $1/L^{2}$, for $T_L$ the temperature 
at the peak of the specific heat $c_V = (\partial/\partial T) u$.
(b) $s$ versus $T$, with the inset showing the good coincidence
of the crossing point.}
\label{fig3}
\end{figure}

\begin{figure}[top]
\centerline{\psfig{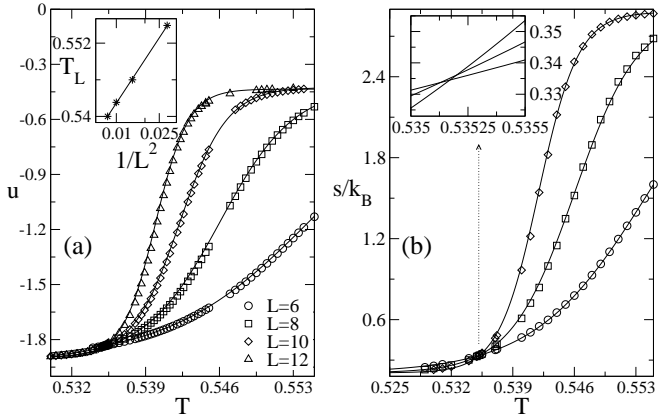}} 
\caption{The same as Fig. 3, but for $q=30$.
For $u$, the curves cross at $T=0.5353(4)$.}
\label{fig4}
\end{figure}

\begin{figure}[top]
\centerline{\psfig{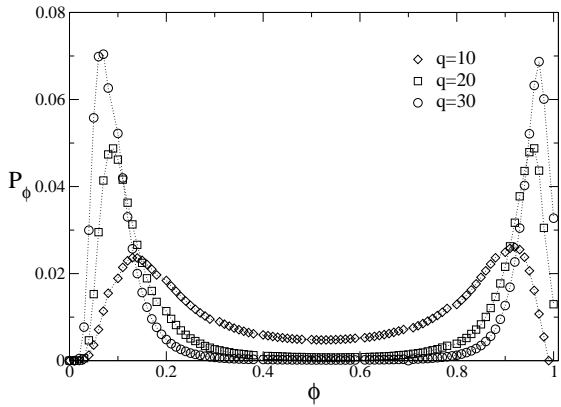}} 
\caption{For different Potts model $q$ values and $L=10$, histograms
of the order parameter probability $P_{\phi}$ versus $\phi$
at the coexistence.}
\label{fig5}
\end{figure}

Finally, in Fig. \ref{fig5} we display histograms of $\phi$ 
for three different values of $q = 10, 20, 30$ at the coexistence.
It illustrates that stronger the phase transition (i.e., higher the $q$'s),
lesser the overlap between the peaks (centered at the distinct phases
$\phi$ values) of the order parameter bimodal distribution.
In particular, observe a larger overlap for $q=10$, for which we find
from the present method a transition at $T_0 = 0.7016(9)$ (details not 
shown).
The exact value is $1/\ln[1+\sqrt{10}] = 0.701231(\ldots)$.
Therefore, although still good, it is not so accurate as the previous
examples.

\subsection{Bell-Lavis model}

The Bell-Lavis (BL) \cite{bell} is a lattice gas model able to reproduce
liquid polimorphism and water-like anomalies. 
It is defined on a triangular lattice where each site is characterized
by its occupation ($\sigma$) and orientation ($\tau$) states. 
Whether the site $i$ is or is not occupied by a water molecule, 
$\sigma_i=1$ or $\sigma_i=0$, respectively.
Furthermore, if the site $i$ has an ``arm'' which is (is not) inert 
towards the adjacent site $j$, then $\tau_{i}^{i j} = 0$ 
($\tau_{i}^{i j} = 1$).

\begin{figure}[top]
\centerline{\psfig{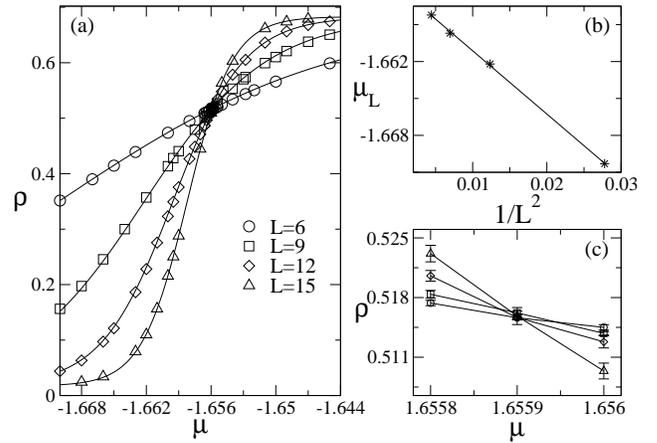}}
\caption{For the BL model, 
(a) $\rho$ versus $\mu$ for $T=0.3$ and distinct $L$'s around the gas-LDL 
transition.                
All the continuous lines are properly obtained from Eq. (\ref{eq1}).  
(b) The values of $\mu=\mu_L$ (for which $(\partial/\partial \mu) \rho$
is maximum) versus $1/L^{2}$. 
(c) A blow up of (a) around the crossing point and the respective small
error bars.}
\label{fig6}
\end{figure}

Two nearest neighbor molecules $i$ and $j$ interact via a van-der-Waals 
energy $-\epsilon_{vdw}$.
They also form a hydrogen bond (of energy $-\epsilon_{hb}$) when 
$\tau_{i}^{ij}\tau_{j}^{ji} = 1$.
So, in the grand-canonical ensemble the BL is described by the 
Hamiltonian
\begin{equation}                                                               
{\mathcal H} = -\sum_{<i,j>} \sigma_{i} \, \sigma_{j} \,                       
(\epsilon_{hb} \, \tau_{i}^{ij} \, \tau_{j}^{ji} +                             
\epsilon_{vdw}) - \mu \sum_{i} \sigma_{i}.                                    
\label{hambl}                                                                  
\end{equation}
The van-der-Waals interaction favors an increasing in the lattice
density (a proper order parameter), whereas the hydrogen bond tends to 
form sublattices for which the molecules have opposite orientations. 

For $\zeta = \epsilon_{vdw}/\epsilon_{hb} < 1/3$, the system  presents three 
stable phases, named: gas, low-density-liquid (LDL), and high-density-liquid 
(HDL). 
For low $\mu$, the system is in the gas phase. 
By increasing $\mu$ we go through the gas-LDL and then through the 
LDL-HDL phase transitions.
At zero temperature both gas-LDL and LDL-HDL are of first order, taking
place at $\mu_c = -3 \, (1+\zeta)/2$ and $\mu_c = -6 \, \zeta$, 
respectively.
For $T > 0$, the gas-LDL remains first-order (ending up in a tricritical 
point if $\zeta=1/10$), but the LDL-HDL becomes second-order 
\cite{bellfiore,fiore-luz2,fiore2011,szortyka}.

For  $T=0.3$, in Fig. \ref{fig6} (a) we plot $\rho$ versus $\mu$ around 
the gas-LDL coexistence.
As previously, the isotherms are well described by Eq. (\ref{eq1}),
with a crossing occurring at $\mu_0=-1.6559(1)$ for
$\rho \approx 0.516(2)$.
Such $\rho$ is close to $1/2$, the exact result at $T=0$ (understood 
recalling that at the coexistence both gas ($\rho=0$) and LDL ($\rho = 2/3$) 
phases have equal weight and the LDL has degeneracy $\alpha_{LDL} = 3$).
In Fig. \ref{fig6} (b) we display $\mu_L$ (the $\mu$ for which 
$(\partial/\partial \mu) \rho$ is a maximum) versus $1/L^{2}$. 
By taking the thermodynamic limit $L \rightarrow \infty$, we find
the extrapolated value of $\mu_0=-1.6560(1)$, in excellent agreement 
with the estimate in Fig. \ref{fig6} (a).
Finally, Fig. \ref{fig6} (c) is a blow up of Fig. \ref{fig6} (a) in the 
vicinity of the phase coexistence.
The observed small error bars illustrate the good accuracy of the present 
method in locating the transition point.

We should stress that although this is the only instance where we present 
a more detailed error analysis, in all the other examples the error bars are 
likewise small.

\subsection{Associating lattice-gas (ALG) model}

Similarly to the BL, the symmetric associating lattice-gas (ALG) model 
\cite{alg1,alg2} can display liquid polimorphism and water anomalies.
It is also defined on a triangular lattice, where
each site is described by an occupation ($\sigma$) and orientation
($\tau$) state. 
But an important difference from the BL is that an energetic punishment
exists when a hydrogen bond is not formed.   
Two first neighbor molecules have an interaction energy of $-v$
($-v + 2 u$) if there is (there is not) a hydrogen bond between them.
The Hamiltonian therefore reads
\begin{equation}                                                               
{\cal H} = 2 u \sum_{<i,j>} \sigma_{i} \sigma_{j} [(1 - v/(2 u)) -             
\tau_{i}^{ij} \tau_{j}^{ji}] - \mu \sum_{i}\sigma_i.                           
\end{equation}

The ALG presents a gas and two liquid, LDL and HDL, phases. 
In particular, for the LDL phase $3/4$ of the lattice is filled by water 
molecules forming the maximum number of hydrogen bonds \cite{alg2}.  
Another relevant distinction from the BL model is that here both
gas-LDL and LDL-HDL transitions remain first-order for $T \neq 0$. 
At $T=0$, the discontinuous transitions occur at $\mu/v=-2$ 
(gas-LDL) and $\mu/v=-6+8u/v$ (LDL-HDL). 

Around the transitions gas-LDL, Fig. \ref{fig7} (a), and LDL-HDL, 
Fig. \ref{fig8} (a), we plot the order parameter versus $\mu$ for $T=0.20$ 
and different $L$'s.
For the former, we simple take $\rho$ as the order parameter.
However, since for LDL-HDL the density is never null, we set as the 
order parameter $\phi = (4\rho -3)$.
Both cases are completely described by Eq. (\ref{eq1}), with the
crossing occurring at $\mu_0=-2.0000(2)$, gas-LDL, and $\mu_0=2.0000(2)$, 
LDL-HDL. 
These estimates (within the numerical uncertainties) are identical to 
their exact values at $T=0$, a particularity of the ALG model.
However, by increasing more the temperature, the $\mu_0$'s start to
change as well.
For instance, for gas-LDL (which has a shorter coexistence line than
that for LDL-HDL \cite{alg2}) $\mu_0 = -1.9986(2)$ at $T=0.3$ 
\cite{fioreluzprl}.
On the other hand, for LDL-HDL it is necessary $T > 0.5$ for a sensible 
departure of $\mu_0$ from its value of 2 at $T=0$ (see also Sec. V), 
e.g., for $T = 0.6$, we have found that $\mu_0 = 1.9970(5)$ (results 
not shown).

\begin{figure}[top]
\centerline{\psfig{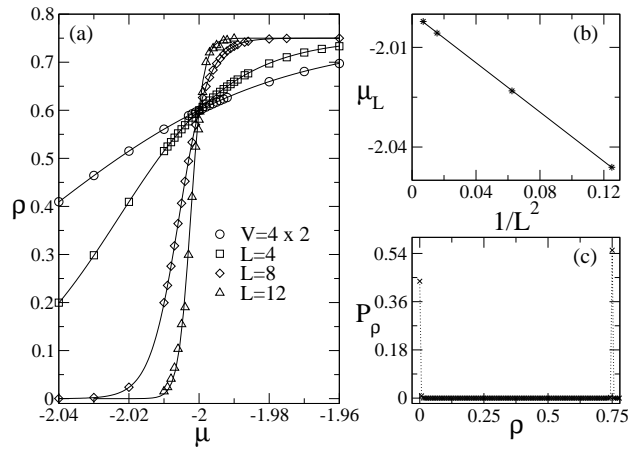}}
\caption{For the ALG model and the gas-LDL transition, (a) the density 
$\rho$ versus $\mu$ for distinct volumes and $T=0.2$, 
(b) the linear scaling of $\mu_{L}$ versus $1/L^{2}$, and 
(c) the probability density $P_\rho$ versus $\rho$ for $L=12$ at 
the coexistence. Results for $V=4 \times 2$ are exact.}
\label{fig7}
\end{figure}

\begin{figure}[top]
\centerline{\psfig{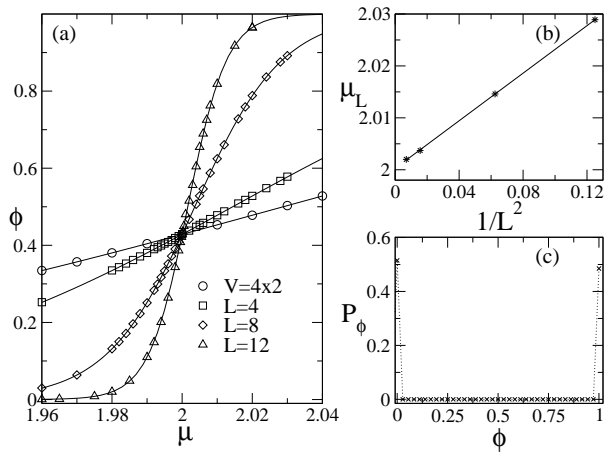}}
\caption{Similar to Fig. \ref{fig7}, but for the LDL-HDL transition
and the order parameter $\phi$ (see main text).}
\label{fig8}
\end{figure}

In Figs. \ref{fig7} and \ref{fig8} (b) we plot $\mu_{L}$ versus $1/L^2$, 
for $\mu_L$ the $\mu$ for which the respective response functions,
$(\partial/\partial \mu) \rho$ and $(\partial/\partial \mu) \phi$,
present a maximum.
Extrapolating to the thermodynamic limit we get the estimates 
$\mu_0=-1.9999(1)$ and $\mu_0=2.0000(1)$, values very close to those
from the crossing calculation.
Lastly, the bimodal density probability distributions, $P_\rho$ and 
$P_\phi$, are shown Figs. \ref{fig7} and \ref{fig8} (c).
They present a very flat valley between the peaks (with each peak
associated to an individual phase at the coexistence).

\subsection{Hard-core gas lattice model}

\begin{figure}[top]
\centerline{\psfig{figure=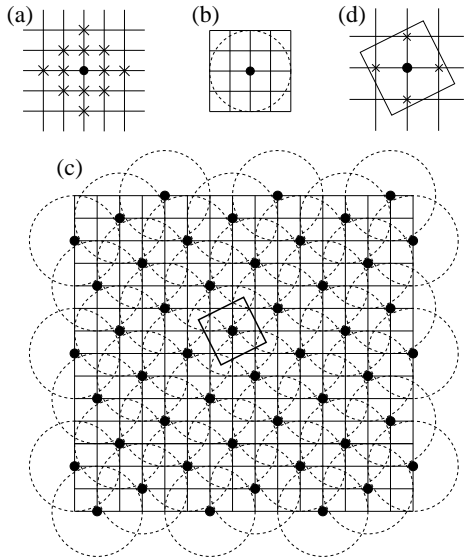,scale=0.3}}
\caption{(a) For the 3NN hard-core gas model, a filled site ($\bullet$) 
hinders neighbor sites ($\times$) to be occupied. 
(b) This corresponds to an spatial exclusion region interaction
(dashed circle).
(c) The regular structure in the maximum possible filling configuration 
has (d) an unitary cell with four vacant sites (in a total of five), 
resulting in a density of $\rho_{max} = 1/5$.}
\label{fig9}
\end{figure}

\begin{figure}[top]
\centerline{\psfig{figure=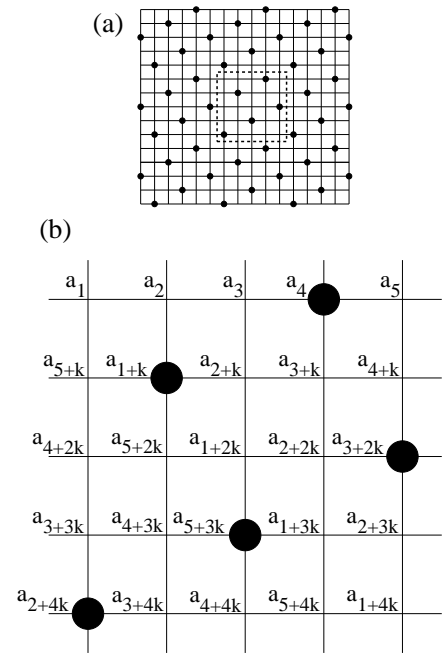,scale=0.30}}
\caption{(a) A basic cell (of $5 \times 5$ sites) for the 3NN hard-core 
gas model (here illustrated in the maximum filling condition).
The full lattice is formed by juxtaposing basic cells.
(b) The basic cell sites labeling, with $a_{i + 5 k} = a_{i}$ for
$i = 1,\ldots,5$ and $k = 0,\ldots,4$.
So, there are five distinct ways to name a basic cell, but once one is 
assumed, it should be used throughout.}
\label{fig10}
\end{figure}

As a last example, let us assume a 2D square lattice for the hard-core gas 
model introduced in \cite{ref11} and recently revisited in 
\cite{ref12,ref13}.  
The interaction is entirely given by an exclusion range: a particle in 
a certain site prevents the occupation of all the surrounding sites.
In the so called 3NN version, the one analyzed here, the excluded sites are 
those shown in Fig. \ref{fig9} (a)-(b).
In this case, the lattice maximum possible filling is displayed in 
Fig. \ref{fig9} (c), resulting in a density of $\rho_{max}=1/5$ 
(Fig. \ref{fig9} (d)). 

The sole control parameter is an effective chemical potential $\tilde{\mu}$, 
which determines the total number of particles in the system.
Hence, temperature is not defined, characterizing an athermal problem.
For a fixed $\tilde{\mu}$, the probability to have $n$ particles
in a lattice of volume $V$ is given by 
$p(n) = \alpha_n \exp[\tilde{\mu} \, n]/Z$,
for $Z = \sum_{n=0}^{n=N_{max}} \alpha_n \exp[\tilde{\mu} \, n]$,
$N_{max} = V/5$, and 
$\alpha_n$ the number of distinct configurations of $n$ particles
allowed by the hard-core potential.
By decreasing (increasing) $\tilde{\mu}$, the system density decreases
(increases).
For lower $\tilde{\mu}$'s, the particles are basically randomly distributed 
-- obeying the above restrictions -- constituting a fluid phase. 
For higher $\tilde{\mu}$'s, the system starts to present a certain ordering
so to accommodate larger numbers of particles (up to a maximum of 
$N_{max}$).
The transition fluid-ordering is of first-order (actually, in the ordering 
regime there are two phases related to each other by a chiral 
transformation \cite{ref12,ref13,ref14}).

Since the usual density $\rho$ only vanishes for 
$\tilde{\mu} \rightarrow -\infty$ and the phase transition takes place
for a finite $\tilde{\mu}$, a more appropriate order parameter $\phi$ 
should be considered.
For so, we follow the approach in \cite{ref13} and take the full
lattice as composed of basic cells (sublattices) of $5 \times 5$ sites
each, Fig. \ref{fig10} (a).
The sites in each row of a basic cell are labeled as $a_i$ ($i=1,\ldots,5$) 
and in total there are five different ways ($k=0,\ldots,4$) to name it 
(see Fig. \ref{fig10} (b)).
Thus, from such construction one can set $\phi$ as in \cite{ref13}
(just using a slight different notation), or 
\begin{equation}
\phi = \langle |\phi_{k=3}-\phi_{k=0}| \rangle,
\ \ \phi_{k} =  \frac{\rho_{max}}{4} \sum_{i,j=1;j>i}^{5} |n_{a_i}^k-n_{a_j}^k|.
\label{op-hard-core}
\end{equation}
Above, $\langle \ldots \rangle$ denotes average over all the lattice
basic cells.
Also, for a chosen labeling $k$, $n_{a_i}^k$ denotes the number 
of particles in the sites named $a_i$ of a basic cell.
According to this definition, at the maximum filling $\phi_{k=0}=0$ and 
$\phi_{k=3}=1$, so $\phi = 1$.
On the other hand, at low densities (fluid phase) $\phi_{k}$ tends to 
zero, consequently it does $\phi$.

\begin{figure}[top]
\centerline{\psfig{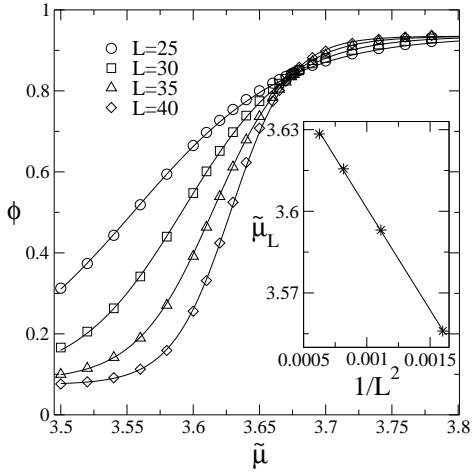}}
\caption{For the 3NN hard-core gas model, the order parameter 
$\phi$ versus the effective chemical potential $\tilde{\mu}$ for 
different $L$'s. 
Continuous lines are from Eq. (\ref{eq1}). 
The curves cross at $\tilde{\mu}_0=3.6741(8)$. 
In the inset $\tilde{\mu}_L$ versus $1/L^2$, where $\tilde{\mu}_L$
is the position of the peak of the ``susceptibility'' 
$\chi = (\partial/\partial \tilde{\mu}) \phi$.}
\label{fig11}
\end{figure}

The numerical simulation procedure is essentially that described in 
Sec. II.B.
The only small differences are: 
(i) instead to define the replicas at distinct temperatures, they are 
defined at distinct $\tilde{\mu}$'s;
(ii) the occupation state of a site (observing the exclusion rule) 
is changed according to $\min \{1, \exp[\pm \tilde{\mu}]\}$, where
the signal $+$ ($-$) is taken if the site is initially empty (occupied);
finally
(iii) the exchange of configuration between two replicas, say $A$ and 
$B$, is performed according to the probability 
$\min \{1, \, \exp[(\tilde{\mu}_{B} - \tilde{\mu}_{A}) \Delta N]\}$,
with $\Delta N$ the difference of the number of particles of $A$ and
$B$.

We have simulated the model for system sizes ranging from $L=25$ to $L=40$, 
shown in Fig. \ref{fig11}.
Note that all curves are very well described by Eq. (\ref{eq1}), whose 
crossing point occurs at ($\tilde{\mu}_0,\phi_0)=(3.6741(8),0.835(6)$). 
Thus, even for an uncommon (but appropriate) definition for the order
parameter, Eq. (\ref{op-hard-core}), it is properly represented by our 
general function $W$.
In the inset of Fig. \ref{fig11} we plot $\tilde{\mu}_L$ versus $1/L^{2}$, 
where $\tilde{\mu}_L$ is the position of the peak of 
$(\partial/\partial \tilde{\mu}) \phi$.
In the thermodynamic limit we find the value $\tilde{\mu}=3.6758(9)$, 
in agreement with the crossing estimate and with the values $3.6762(1)$
in \cite{ref12} and $3.6746$ in \cite{ref13}. 
In Fig. \ref{fig12} the probability distribution histogram of the order 
parameter for $L=30$ and at the coexistence presents a low valley 
between the peaks of the two coexisting phases.

\begin{figure}[top]
\centerline{\psfig{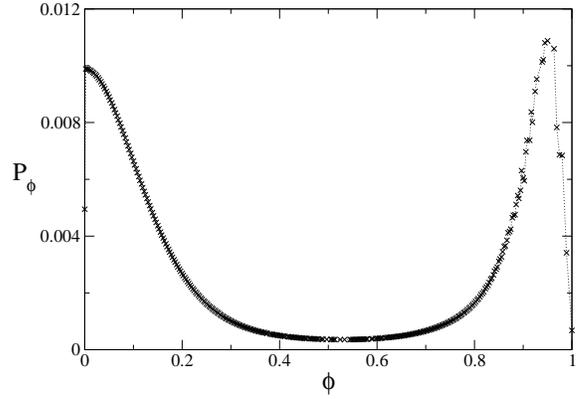}}
\caption{For the 3NN hard-core gas model and $L=30$, histogram of
the order parameter probability $P_\phi$ at the coexistence.}
\label{fig12}
\end{figure}

\section{The method applicability}

So far we have discussed five distinct examples:
a prototype thermodynamic system, three representative lattice-gas 
models, and an interesting athermal problem.
We have found that the present approach is able to describe FOPTs in all 
the different situations studied.
Thus, a relevant issue is to enquire to what extent the method can give good 
results.

To address it, we first recall that the approach key point is the actual 
form of Eq. (\ref{Z-lowT}), i.e., to write $Z$ as a sum of 
exponentials, where each term is uniquely associated to a particular phase. 
In other words, there are no terms involving overlapping between 
coexisting phases.
The rigorous analysis in \cite{borgs1,rBoKo} show that this decomposition
is generally valid at low temperatures (see also the Appendix).

\begin{figure}[top]
\centerline{\psfig{figure=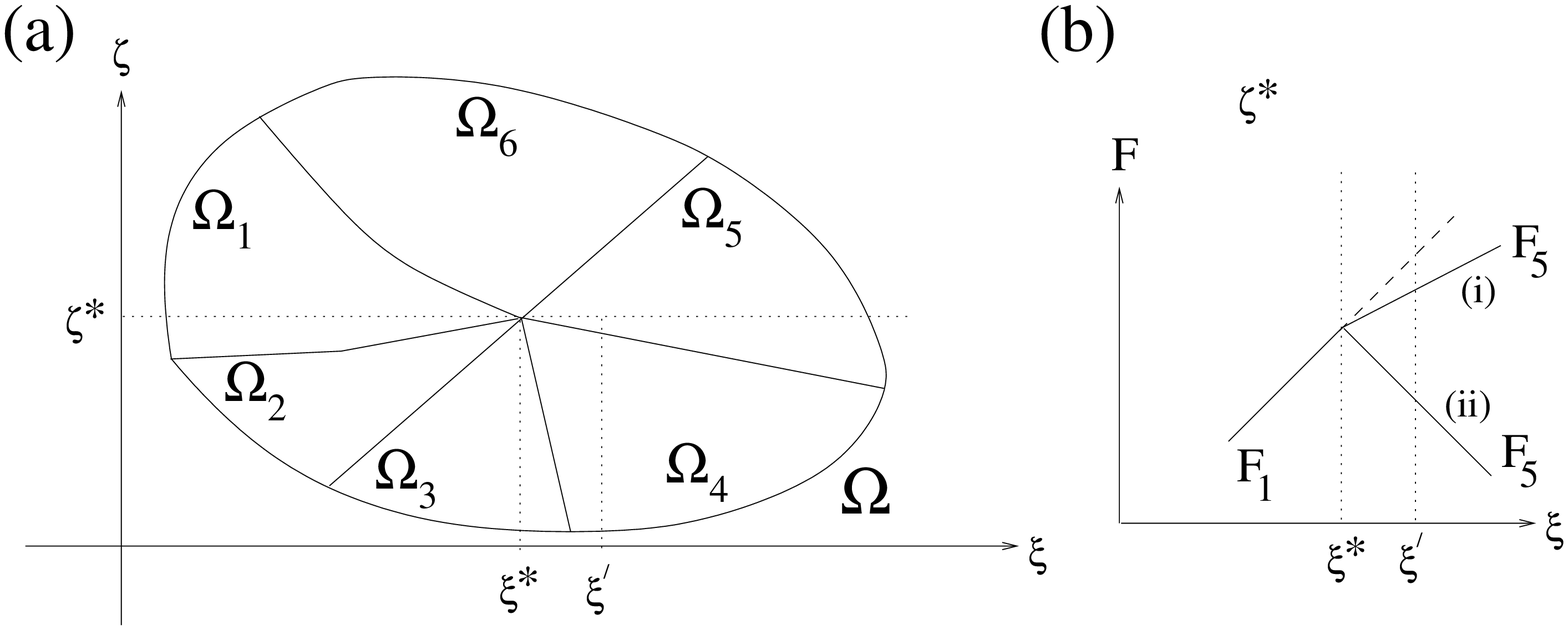,scale=0.3}}
\caption{(a) In an arbitrary phase space $\xi \times \zeta$, 
example of a small region $\Omega$ around a point of coexistence 
of ${\mathcal N}$ phases (here, ${\mathcal N} = 6$).
(b) Schematics of $F$ versus $\xi$ for $\zeta = \zeta^{*}$.
If $\xi > \xi^{*}$, the system is in the phase $n=5$ and the dashed line
represents the functional form of $F_1$ extended into such region.
From the inclination of $F_5$, the FOPT $1$--$5$ would be stronger in 
(ii) than in (i).
So, $\Delta_{1,5}(\xi') = F_1(\xi') - F_5(\xi')$ would be larger in the
case (ii).}
\label{fig13}
\end{figure}

To understand in a more physical ground why of a such structure for the 
partition function, one might consider the following heuristic arguments
(in the specific case we are close to a FOPT point):

\begin{itemize}
\item[(i)]
Assume in phase space a small region $\Omega$ encompassing a point of 
coexistence of ${\mathcal N}$ phases.
Moreover, for any $n = 1,\ldots,{\mathcal N}$, let $\Omega_{n}$ be a portion 
of $\Omega$ corresponding to the phase $n$ (e.g., Fig. \ref{fig13} (a)).
\item[(ii)]
Within any $\Omega_{n}$, $F_{n} = - \ln[Z]/\beta$ is the phase $n$ free 
energy.
Then, suppose (at least formally) that in $\Omega_{n}$ we can write 
$Z \approx \exp[-\beta F_{n}] (1 + \overline{Z}_{n})$, with a proper 
$\overline{Z}_{n}$ being very small in such region 
(in fact, we must have $|\overline{Z}_{n}/\beta| \ll F_{n}$).
\item[(iii)] 
Thus, in each $\Omega_{n}$:  
$Z \approx \exp[-\beta F_{n}] + \mbox{a small term}$.
Therefore, a tentative partition function for the whole $\Omega$ can be 
$Z \approx \sum_{n=1}^{\mathcal N} \exp[- \beta F_n]$
(for simplicity neglecting possible degeneracies $\alpha_n$).
This is just Eq. (\ref{Z-lowT}).
\item[(iv)]
But for the above to hold, a consistency condition is required. 
Notice that $\overline{Z}_{n} = \sum_{n' \neq n} 
\exp[-\beta \Delta_{n',n}]$, with $\Delta_{n', n} = F_{n'} - F_n$ representing the 
difference between the free energy of phases $n'$ and $n$.
So, $\overline{Z}_{n}$ to be small in $\Omega_n$ implies that 
$\forall n' \neq n$, $\beta \Delta_{n',n}$ is large in $\Omega_n$.
\end{itemize}

From the above reasoning we reach the desired general validity 
condition for Eq. (\ref{Z-lowT}) (and so for $W$ in Eq. (\ref{eq1})), 
namely,
\begin{equation}
\beta \Delta_{n',n} \gg 1 \ \mbox{in} \ \Omega_{n} \ \mbox{for any} \
n \ \mbox{and for all} \ n' \neq n.
\label{key-relation}
\end{equation}
Note that it explains why Eq. (\ref{Z-lowT}) is always valid at low $T$'s 
(at least close to phase transitions).
Indeed, in such case even if the $\Delta$'s are not large, the product 
$\beta \Delta$ can be very large if the temperature is sufficiently
small.

But Eq. (\ref{key-relation}) is also true if the $\Delta$'s themselves are 
large (of course, with $T$ not too high).
As illustrated in Fig. \ref{fig13} (b), this is the case in strong FOPTs, 
i.e., for the system displaying large discontinuities in the slop of the 
$F$'s across $\xi^{*}$ or, equivalently, for a large jump in the value of 
the order parameter (for instance, as determined by the quantity $r$ in 
the example of Sec. III).

Lastly, there is a practical and computationally inexpensive test to check 
the above relations.
Equation (\ref{key-relation}) implies in very high entropic barriers 
across the transition point.
Hence, even considering ubiquitous thermodynamic fluctuations (e.g., for 
finite systems) around such point it would be much more probable the order
parameter to assume values typical of the single phases (corresponding 
to the $F$'s minima) than to present values in between (implying the system 
to cross the high $F$'s regions).
So, one could calculate the probability distribution histogram for the 
order parameter at the coexistence condition, for which the peaks relate 
to the distinct phases.
The verification of Eq. (\ref{key-relation}) would result in well separated 
peaks, not overlapping each other.
Indeed, this is exactly the case in the examples here, as observed in Figs. 
\ref{fig1} (right inset), \ref{fig5}, \ref{fig7} (c), 
\ref{fig8} (c), and \ref{fig12}.

\begin{figure}[top]
\centerline{\psfig{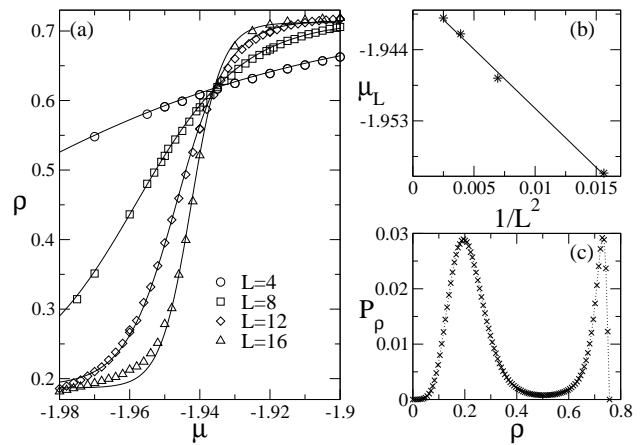}}
\caption{For the ALG model gas-LDL transition at $T=0.5$, 
(a) the density $\rho$ versus $\mu$ for distinct $L$'s.
Continuous lines are obtained from Eq. (\ref{eq1}).
(b) The scaling plot of $\mu_{L}$ (for which 
$\chi = (\langle \rho^2\rangle-\langle \rho \rangle^{2}) V$ is a maximum)
versus $1/L^2$.
(c) The probability density of $\rho$ at the coexistence for $L=12$.}
\label{fig14}
\end{figure}

As a final illustration, we come back to the ALG model of Sec. III.C. 
However, we set $T = 0.5$, which is 2.5 times higher than the value in 
Figs. \ref{fig7} and \ref{fig8}.
This case is interesting because now the gas-LDL (LDL-HDL) transition,
Fig. \ref{fig14} (Fig. \ref{fig15}), is well described by Eq. (\ref{eq1}) 
only closer to the phases coexistence point.
Indeed, compare the fitting quality and the $\mu$ range considered in
Figs. \ref{fig7} and \ref{fig14} and in Figs. \ref{fig8} and \ref{fig15}.
Important to emphasize that unlike Fig. \ref{fig7} (c), in Fig. \ref{fig14} 
(c) the probability distribution is not really a flat valley between the 
peaks for the gas-LDL.
On the other hand, similar to Fig. \ref{fig8} (c), for LDL-HDL the two 
peaks in Fig. \ref{fig15} (c) do not intersect.

Nevertheless, the range in which $W$ is valid is still large enough to allow
a proper characterization of the FOPT in the gas-LDL case (for LDL-HDL, 
the range is even larger, see Fig. \ref{fig15}).
From the crossing curves in Fig. \ref{fig14}, we get the estimate
$\mu_0=-1.9360(5)$, in agreement with $\mu_0=-1.9365(5)$ from the 
position $\mu_L$ of the peak of 
$\chi = (\langle \rho^2\rangle-\langle \rho \rangle^{2}) V$,
Fig. \ref{fig14} (b).
We observe that since the fitting of $\phi$ is not so good for a broader 
interval of $\mu$'s, we prefer to calculate 
$\chi = (\langle \rho^2\rangle-\langle \rho \rangle^{2}) V$ (thus numerically 
more reliable) instead to define $\chi = (\partial/\partial \mu) \phi$.
For the LDL-HDL case, the curves crossing leads to the estimate
$\mu_0=1.9995(5)$, very close to the estimate $\mu_0=1.9990(5)$, 
obtained from the peak of $\chi = (\partial/\partial \mu) \phi$.

\begin{figure}[top]
\centerline{\psfig{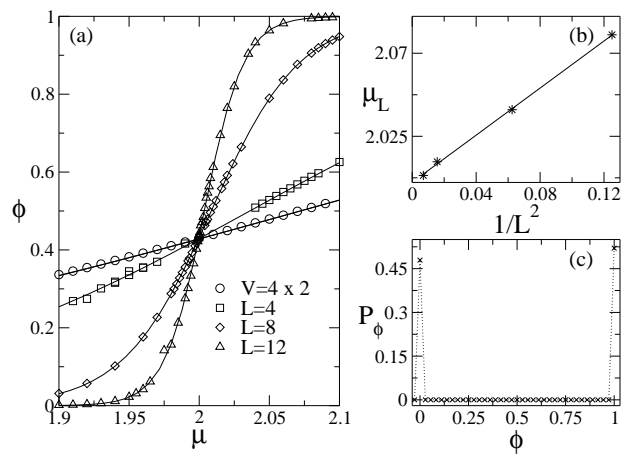}}
\caption{The same as in Fig. \ref{fig14}, but for the LDL-HDL transition.  
Here $\mu_{L}$ is the peak position of 
$\chi = (\partial/\partial \mu) \phi$.}
\label{fig15}
\end{figure}

\section{Remarks and Conclusion}

In this contribution we have clarified the main mathematical aspects, 
discussed the applicability condition and extended the instances of
usage of a recent proposed \cite{fioreluzprl} approach to treat FOPTs,
which can be summarized as the following. 
From a special decomposition for the partition function $Z$, 
Eq. (\ref{Z-lowT}) -- valid close to a FOPT whenever Eq. 
(\ref{key-relation}) holds -- one can derive an analytical 
expression $W$, Eq. (\ref{eq1}), which depends on few coefficients.
By using simple numerical simulations to determine these coefficients, 
$W$ is able to describe quite well relevant thermodynamic quantities,
like order parameter, energy, entropy, etc, around the transition point.
In addition, there is a point where all curves $W$ (irrespect of $L$) 
cross.
By considering relatively small system sizes $L$, the crossing can
give the transition point thermodynamic value.

As it should be, the procedure agrees with other efficient schemes 
available \cite{pottsreview,rBoKo,yang-lee,ch-wang,ref13}.
However, it has the advantage of being general, inexpensive from 
the computational point of view and to yield response functions 
$\chi$ (e.g., compressibility and specific heat) in a rather direct way 
(analytically).

The method validity condition can be tested from plots of multimodal 
probability distributions of the order parameter at the coexistence,
calculated from straightforward simulations.
A non-overlapping of the peaks (associated to the individual phases)
indicates that it can be satisfactorily applied.
In fact, such condition extends the protocol, originally derived 
\cite{fioreluzprl} for the situation of low temperatures.

The method has been tested and shown to work fine for several lattice 
problems, including an exact solvable and the relevant Potts, Bell-Lavis, 
ALG, and athermal hard-core gas, models.
But certainly, a natural question is if FOPTs taking place in off-lattice
systems -- with much larger phase spaces -- could be studied in a similar
fashion.
In this respect, we first observe that some aspects of continuous 
systems displaying FOPTs at low $T$'s, (e.g., as for polymers in Ref. 
\cite{jcp-119-2003}, analyzable in terms of finite-size scaling 
\cite{epl-70-2005}) can be described by lattice models \cite{jcp-131-2009}.
Obviously, in such cases the method could be directly applied.
Second, note that the decomposition for the partition function $Z$, 
when valid, does not make restrictions regarding lattice or 
off-lattice systems. 
Thus, the only issue would be the use (in a continuous phase space) of a 
proper simulation sampling procedure to fit the parameters in Eq. (\ref{eq1}).
For instance, there are some implementations of the PT for off-lattices
systems (see, e.g., Refs. \cite{continuous}).
With such implementations the approach should hold in the same way.

The study of off-lattice \cite{continuous} and polymers systems 
displaying FOPTs \cite{zhou} is presently an ongoing work and will be 
reported in the due course.

\section{Acknowledgments}
We acknowledge research grants from CNPq and Finep-CTInfra.

\appendix

\section{The derivation of $W$ and some of its properties}

\begin{figure}[*t]
\centerline{\psfig{figure=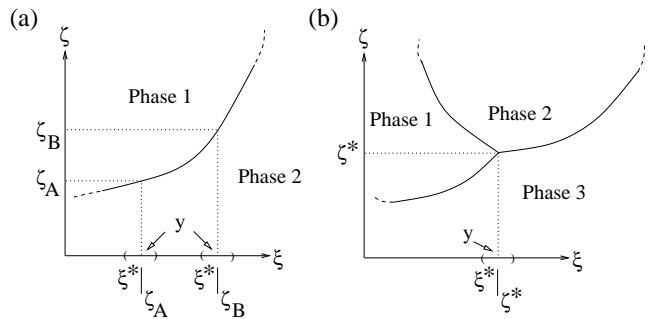,scale=0.39}} 
\caption{Examples of possible phase diagrams.
By varying the control parameter $\xi$ in an interval 
$y = \xi - \xi^{*}$, it is assumed that other intensive quantities, 
$\zeta$, are properly fixed. 
(a) Along the displayed separation line between two phases, 
any $\zeta$ always will allow a $\xi^{*}$. 
(b) On the other hand, for the shown triple point, 
$\zeta$ should be set to $\zeta^{*}$.}
\label{fig:appendix}
\end{figure}

Consider a finite system at a low temperature and presenting ${\mathcal N}$ 
coexisting stable phases (the meaning of ``low'' here is discussed in 
Section V).
It has been rigorously shown \cite{borgs1,rBoKo} that the problem
partition function is well described by (with $\beta = (k T)^{-1}$
and $V$ the volume)
\begin{equation}
Z = \sum_{n=1}^{\mathcal{N}} \, \alpha_n \, \exp[-\beta V f_n] + Z_{unst}.
\label{a1}
\end{equation}
$Z_{unst}$ is associated to the possible existence of unstable phases
-- but which are exponentially damped, so negligible in Eq. (\ref{a1}) -- 
and $f_n$ is the $n$-th phase ($n = 1, \ldots, \mathcal{N}$) free energy 
per volume \cite{rBoKo}.
The degeneracy parameters (or weights) $\alpha$'s result from eventual 
symmetries, so that $\alpha_n > 1$ would be due to distinct spatial 
configurations leading to a same phase $n$.

Now, let $\xi$ to be a proper phase transition control parameter,
which we shall vary.
From Eq. (\ref{a1}) we define 
\begin{equation}
W = -\frac{1}{\beta \, V} \frac{\partial}{\partial \xi} 
\ln[Z]
= 
\frac{\sum_{n=1}^{\mathcal{N}} \, \alpha_n \, g_n \exp[-\beta V f_n]}
{\sum_{n=1}^{\mathcal{N}} \, \alpha_n \exp[-\beta V f_n]},
\label{a2}
\end{equation}
with 
\begin{equation}
g_n = T \frac{\partial}{\partial \xi} \left(\frac{f_n}{T}\right).
\end{equation}
Note this the definition of $W$ is usually the start point to 
calculate relevant order parameters.

The system may have many other intensive parameters, which
we generally denote by $\zeta$.
So, suppose $\zeta$ kept fixed at proper values, such that the 
$\mathcal{N}$ phases coexistence takes place at $\xi = \xi^{*}$
(usually with $\xi^{*}$ depending on $\zeta$, see Fig. 
\ref{fig:appendix}).
In this case, the $f_n$'s are single functions of $\xi$ and for 
$y = \xi - \xi^{*} \approx 0$ we consider a first order series 
expansion (possible due to the existence of smooth representations 
for the $f_n$'s \cite{borgs1}):
$f_n \approx f^{*} + f_n^{\prime *} \, y$, with
$f_n(\xi^{*}) = f^{*}$ $\forall$ $n$ (because the coexistence) and
$f_n^{\prime *} = (\partial f_n/\partial \xi)|_{\xi=\xi^{*}}$.
It leads to
\begin{equation}
g_n \approx f_n^{\prime *} - \frac{f^*}{T^*} \frac{\partial T}{\partial \xi}, 
\label{gn}
\end{equation}
where, obviously, $\partial T/\partial \xi = 1$ for $\xi = T$ and zero 
otherwise.
In deriving Eq. (\ref{gn}), if the control parameter $\xi$ is the 
temperature, one also must assume $1/T \approx 1/T^*$, a reasonable 
approximation provided $W$ is calculated for $|T - T^*|$ small 
(for how small in practice, see the numerical examples along the paper).

Next, for 
$a_n = (f_n^{\prime *} - f_1^{\prime *}) \, V \beta$
[or $a_n = (f_n^{\prime *} - f_1^{\prime *}) \, V \beta^*$ if $\xi = T$],
$b_n = (\alpha_n/\alpha_1) \, g_n$, and
$c_n = (\alpha_n/\alpha_1)$, we get
\begin{equation}
W \approx \frac{b_1 + \sum_{n=2}^{\mathcal{N}} b_n \, \exp[-a_n y]}
       {1 + \sum_{n=2}^{\mathcal{N}} c_n \, \exp[-a_n y]}.
\label{a3}
\end{equation}
Above, the coefficients $a_n$, $b_n$ and $c_n$ are independent on the
control parameter and only the $c_n$'s depend (linearly) on the
volume. 
Hence, at the coexistence ($y=0$) $W \neq W(V)$ and all the curves 
$W$ versus $\xi$, regardless of $V$, must cross at $\xi=\xi^{*}$.
Thus, Eq. (\ref{a3}) not only describes generic thermodynamic 
quantities, but also yields the thermodynamic limit estimate for the 
transition point.
Furthermore, by taking derivatives of Eq. (\ref{a3}), one can obtain
response functions and susceptibilities.

Finally, at $\xi = \xi^{*}$ either from Eq. (\ref{a2}) or from Eq. 
(\ref{a3}) $W$ reads
\begin{equation}
W(\xi^{*}) = \sum_{n=1}^{\mathcal{N}} p_n f_n^{\prime *} -
\frac{f^*}{T^*} \frac{\partial T}{\partial \xi},
\label{a4}
\end{equation}
with
\begin{equation}
p_n = \frac{\alpha_n}{\sum_{n=1}^{n=\mathcal{N}} \alpha_n}.
\end{equation}
Moreover, suppose the $f$'s ordered such that
$f_{-}^{\prime *} = f_1^{\prime *} = \ldots  = f_m^{\prime *} 
< f_{m+1}^{\prime *} \leq \ldots \leq f_{k-1}^{\prime *} < f_{k}^{\prime *} = 
\ldots = f_{\mathcal{N}}^{\prime *} = f_{+}^{\prime *}$.
Assuming $y$ small, so we can consider Eq. (\ref{a3}), if we take
$V\rightarrow \infty$ with $y$ positive (case $+$) or $y$ negative
(case $-$), for $v_+ = 1, u_+ = m, v_- = k, u_- = {\mathcal N}$
we find (note a little misprint in Ref. \cite{fioreluzprl})
\begin{equation}
W_{\pm} = \frac{\sum_{n=v_{\pm}}^{n=u_{\pm}} b_n}{\sum_{n=v_{\pm}}^{n=u_{\pm}} c_n}
=
f_{\mp}^{\prime *} -
\frac{f^*}{T^*} \frac{\partial T}{\partial \xi}.
\label{a5}
\end{equation}
Equations (\ref{a4})-(\ref{a5}) give the discontinuity of the
order parameter across the phase transition in the thermodynamic
limit.
Numerically, such discontinuity is obtained from the coefficients 
$b$'s and $c$'s once, from Eq. (\ref{a5}), we can write $W_{+} = b_1/c_1$ 
and $W_{-} = b_{\mathcal N}/c_{\mathcal N}$.
In particular, for $k={\mathcal N}$ ($m=1$), $W_{+}$ ($W_{-}$) is 
given in terms of the sole phase which is immediately to the right 
(left) of $\xi^{*}$.
Also, the number of equal $a_n$'s correspond, at least in a first order 
approximation, to the number of phases which coexist over the line $\xi$
in the vicinity of $\xi^{*}$.
This fact can be used to locate coexisting phase lines and triple points
(an application for the present method to appear elsewhere).

\end{document}